\documentclass[pre,twocolumn,showpacs,amsmath,amssymb,floatfix]{revtex4-1}
\usepackage{graphicx}
\usepackage{bm}

\newcommand{\f}{\frac}

\def\a{\alpha}
\def\bta{\beta}

\def\la{\langle}
\def\ra{\rangle}

\def\g{\gamma}

\def\bx{{\bf x}}

\begin{document}

\title{Application of importance sampling to the computation of large
  deviations in non-equilibrium processes}


\author{Anupam Kundu, Sanjib Sabhapandit, and Abhishek Dhar}
\affiliation{Raman Research Institute, Bangalore 560080, India}
\date{\today}

\begin{abstract}
We present an algorithm for finding the probabilities of rare events in
nonequilibrium processes. The algorithm consists of evolving the system
with a modified dynamics for which the required event occurs more
frequently. By keeping track of the relative weight of phase-space
trajectories generated by the modified and the original dynamics one can
obtain the required probabilities.  The algorithm is tested on two model
systems of steady-state particle and heat transport where we find a huge
improvement from direct simulation methods.
\end{abstract}

\pacs{05.40.--a,05.10.Ln,05.70.Ln} 
\maketitle

\section{Introduction}

A rare event is one which occurs with a very small probability. However,
when they do occur they can have a huge effect and so it is often important
to estimate the actual probability of their occurrence.  Examples where
rare events are important are in banking and insurance, in biological
systems where important processes such as genetic switching and mutations
occur with extremely small rates, and in nucleation processes. Rare events
are also of importance in nonequilibrium processes such as charge and heat
transport in small devices and transport in biological cells.  The
functioning of nano-electronic devices can be affected by rare
large-current fluctuations and it is important to know how often they
occur.

In this paper our interest is in predicting probabilities of rare
fluctuations in transport processes.  A number of interesting results have
been obtained recently on large fluctuations away from typical behavior in
nonequilibrium systems.  These include results such as the fluctuation
theorems~\cite{ECM93,JW04,BD04,DDR04,DL98,visco06,SD07} and the Jarzynski
relation \cite{jarz97}.  In the context of transport one typically
considers an observable, say $Q$, such as the total number of particles or
heat transferred across an object with an applied chemical potential or
temperature difference. This is a stochastic variable and for a given
observation time $\tau$ this will have a distribution $P(Q,\tau)$. The
various general results that have been obtained for $P(Q,\tau)$ give some
quantitative measure of the probability of rare fluctuations. Analytic
computations of the tails of $P(Q,\tau)$ for any system are usually
difficult.  This is also true in experiments or in computer simulations
since the generation of rare events requires a large number of trials.

For large $\tau$ the probabilities of large fluctuations show scaling
behavior $P(Q,\tau) \sim e^{-\tau\,f(Q/\tau)}$, where the function $f(q)$
is known as the large deviation function~\cite{Varadhan, Touchette}.  For a
few model systems exact results have been obtained~\cite{DL98,visco06,SD07}
for either $f(q)$ or its Legendre transform $\mu(\lambda)$, which can be
defined in terms of the characteristic function as
$\mu(\lambda)=\lim_{\tau\rightarrow\infty}\tau^{-1}\ln \bigl\langle
e^{-\lambda Q}\bigr\rangle$.  Recently an algorithm has been proposed
\cite{GKP06} to compute $\mu(\lambda)$. However, as has been pointed out in
Ref.~\cite{HG09} there may be problems in obtaining the tails of
$\mu(\lambda)$ using the algorithm of Ref.~\cite{GKP06}. The algorithm
proposed in this paper is complementary to the one discussed in
Ref.~\cite{GKP06} in the sense that we obtain $P(Q,\tau)$ directly. Our
algorithm, based on the idea of importance sampling, computes $P(Q,\tau)$
for any given $\tau$ and accurately reproduces the tails of the
distribution.  Algorithms based on importance sampling \cite{Bucklew} have
earlier been used in the study of equilibrium systems
\cite{rosenbluth,Grassberger:97} and in the study of transition rate
processes \cite{MJB98,dellago98,allen05}.  However, we are not aware of any
applications to the study of large fluctuations of currents in
nonequilibrium systems and this is the main focus of this paper. Here we
choose two prototype models of transport, namely, heat conduction across a
harmonic chain and particle transport in the symmetric simple exclusion
process.  We illustrate the implementation of importance sampling in the
computation of large fluctuations of currents in these two nonequilibrium
systems.
 
Consider a system with a time evolution described by the stochastic process
$x(t)$. For simplicity we assume for now that $x(t)$ is an integer-valued
variable and time is discrete. Let us denote a particular path in
configuration space over a time period $\tau$ by the vector $\bx(\tau) :=
\{x(t)| t=1,2,\dotsc,\tau\}$ and let $Q$ be an observable which is a
function of the path $\bx(\tau)$. We will be interested in finding the
probability distribution $P(Q,\tau)$ of $Q$ and especially in computing the
probability of large deviations about the mean value $\la Q \ra$.  As a
simple illustrative example consider the tossing of a fair coin. For
$\tau=N$ tosses we have a discrete stochastic process described by the time
series $\bx(N) = \{x_i\}$ where $x_i=1$ if the outcome in the $i$th trial
is heads and $x_i=-1$ otherwise.  Suppose we want to find the probability
of generating $Q$ heads (thus $Q=\sum_{i=1}^N \delta_{x_i,1}$). An example
of a rare event is, for example, the event $Q=N$.  The probability of this
is $2^{-N}$ and if we were to simulate the coin toss experiment we would
need more than $2^N$ repeats of the experiment to realize this event with
sufficient frequency to calculate the probability reliably. For large $N$
this is clearly very difficult.  The importance sampling algorithm is
useful in such situations. The basic idea is to increase the occurrence of
the rare events by introducing a bias in the dynamics. The rare events are
produced with a new probability corresponding to the bias. However, by
keeping track of the relative weights of trajectories of the unbiased and
biased processes it is possible to recover the required probability
corresponding to the required unbiased process.

\section{The algorithm}

We now describe the algorithm in the context of
evaluating $P(Q,\tau)$ for the stochastic process $\bx(\tau)$. We denote
the probability of a particular trajectory by ${\mathcal{P}}(\bx)$ . By
definition:
\begin{equation} 
P(Q,\tau)= \sum_{\bx} \delta_{Q,Q(\bx)} \mathcal{P}(\bx).
\end{equation} 
For the same system let us consider a biased dynamics for which the
probability of the same path $\bx$ is given by ${\mathcal{P}}_b(\bx)$. Then
we have:
\begin{eqnarray} 
P(Q,\tau) &=& \sum_{\bx} \delta_{Q,Q(\bx)} e^{-W(\bx)}
\mathcal{P}_b(\bx),  \\
\text{where}\quad e^{-W(\bx)} &=&
\frac{\mathcal{P}(\bx)}{\mathcal{P}_b(\bx)}.  
\end{eqnarray} 
Thus in terms of the biased dynamics, $P(Q,\tau)$ is the average $\la
\delta_{Q,Q(\bx)} e^{-W} \ra_b $ and in a simulation we estimate this by
performing averages over $M$ realizations to obtain:
\begin{equation} P_e(Q,\tau)=\frac{1}{M} \sum_r
   {\delta_{Q,Q({\bx_r})} e^{-W(\bx_r)}} ~, \label{Pest}
\end{equation} 
where $\bx_r$ denotes the path for the $r$th realization.  For $M \to
\infty$ we obtain $P_e(Q,\tau) \to P(Q,\tau)$ which is the required
probability.  Note that the weight factor $W$ is a function of the path. In
a simulation we know the details of the microscopic dynamics for both the
biased and unbiased processes. Thus we can evaluate $W$ for every path
$\bx$ generated by the biased dynamics. A necessary requirement of the
biased dynamics is that the distribution of $Q$ that it produces [i.e.,
  $P_b(Q,\tau)=\langle \delta_{Q,Q(\bx)}\rangle_b$] should be peaked around
the desired values of $Q$ for which we want an accurate measurement of
$P(Q,\tau)$. As we will see the required dynamics can often be guessed from
physical considerations.

We first explain the algorithm for the coin tossing experiment.  In this
case we consider a biased dynamics where the probability of heads is $p$
and that of tails is $1-p$. If we take $p \approx 1$ then the event $Q=N$,
which was earlier rare, is now generated with increased frequency and we
can use Eq.~(\ref{Pest}) to estimate the required probability
$P(Q=N,N)$. For any path consisting of $Q$ heads the weight factor is
simply given by $e^{-W}=(1/2)^{N}/[p^Q(1-p)^{N-Q}]$. Choosing $p=0.95$ it
is easy to see that for $N=100$ we can get the required probability
$P(Q=N,N)$ with more than $1 \%$ accuracy using only $M=10^7$ realizations
as opposed to at least $M=10^{30}$ required by the unbiased dynamics.  Note
that for this example $W$ has the same value for all paths with the same
$Q$.  In general of course $W$ depends on the details of the path,
{\emph{e.g.}} for a random walk with a waiting probability.  We will now
illustrate the algorithm with non-trivial examples of computing large
deviations in two well known models in nonequilibrium physics. These are
the (i) symmetric simple exclusion process (SSEP) with open boundaries and
(ii) heat conduction across a harmonic system connected to Langevin
reservoirs.
 
\section{Symmetric simple exclusion process}

This is a well studied example of an interacting stochastic system
consisting of particles diffusing on a lattice with the constraint that
each site can have at most one particle. Here we restrict ourselves to
one-dimension and study the case of an open system where a linear chain
with $L$ sites is connected to particle reservoirs at the two ends.  The
dynamics can be specified by the following rules: (a) a particle at any
site $l=1,2,\dotsc,L$ can jump to a neighboring empty site with unit rate;
(b) at $l=1$ a particle can enter the system with rate $\a$ (if it is
empty) and leave with rate $\g$. At site $N$ a particle can leave or enter
the system with rates $\beta$ and $\delta$, respectively.  The biased
dynamics can be realized in various ways, for example, by introducing
asymmetry in the bulk hopping rates or by changing the boundary hopping
rates.

\begin{figure}
\includegraphics[width=3.25in]{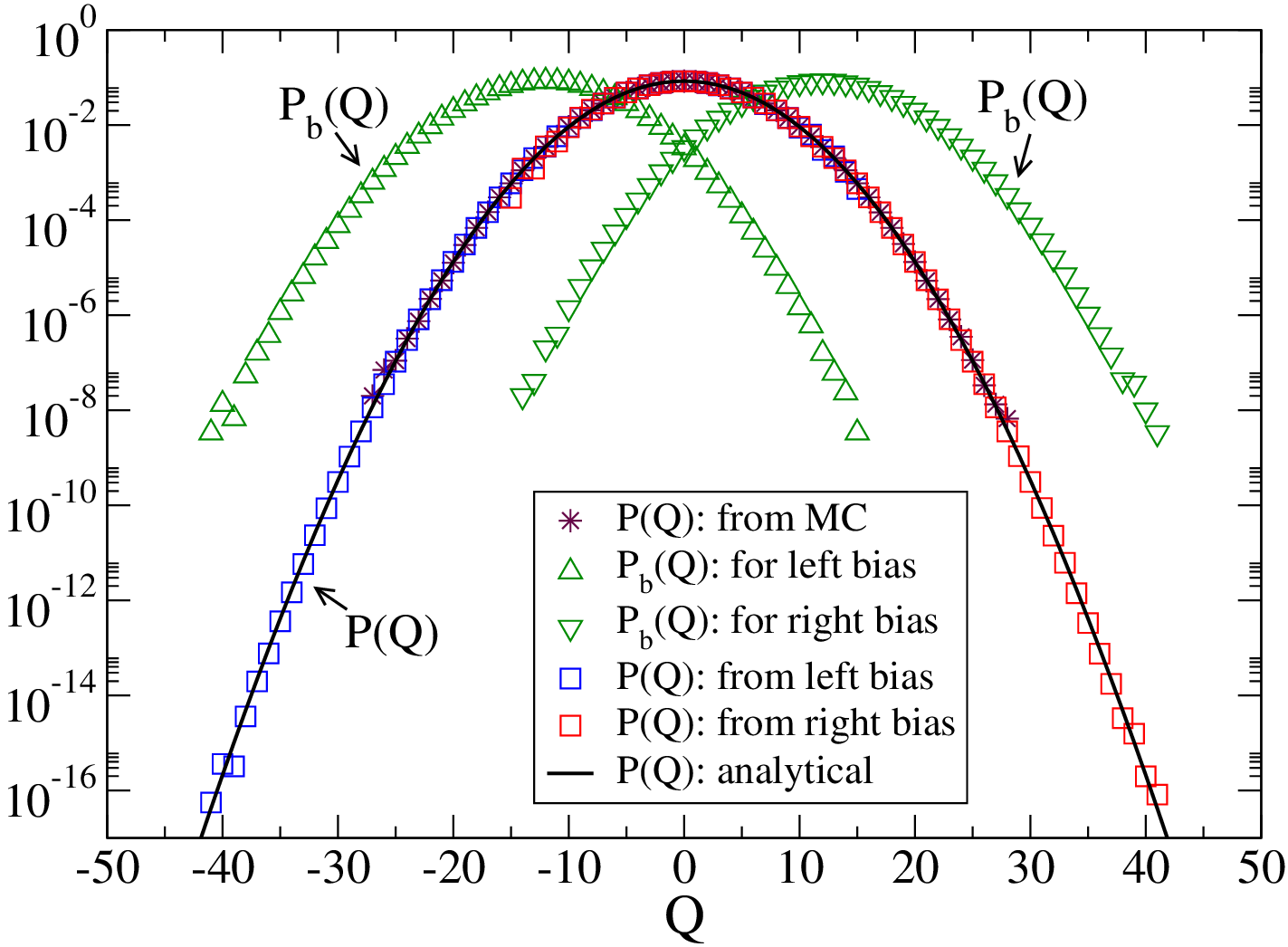}
\caption{(Color online) Plot of $P(Q)$ for $\tau=15$ for the one-site SSEP
  model with $\a=\bta=3.0,\g=\delta=3.0$. MC refers to direct Monte Carlo
  simulations.  Left bias corresponds to $\a'=\bta'=3.8,\g'=\delta'=2.2$
  and right bias to $\a'=\bta'=2.2,\g'=\delta'=3.8$.}
\label{fig-1-part}
\end{figure}

For SSEP, the configuration of the system at any time is specified by the
set ${\mathcal{C}}=\{n_1(t),n_1(t),...,n_L(t)\}$ where $n_l(t)$ ($0$ or
$1$) gives the occupancy of the $l$th site. The dynamical rules specify the
matrix element $\mathcal{W}({\mathcal{C}},{\mathcal{C'}})$ giving the
transition rate from configuration ${\mathcal{C}}'$ to ${\mathcal{C}}$. We
write $\mathcal{W}
({\mathcal{C}},{\mathcal{C'}})=\mathcal{W}_1+\mathcal{W}_{-1}+\mathcal{W}_0$
where $\mathcal{W}_1$ and $\mathcal{W}_{-1}$ correspond to transitions
whereby a particle enters the system from the left bath or leaves the
system into the left bath, respectively, while $\mathcal{W}_0$ corresponds
to all other transitions.  At long times the system will reach a steady
state with particles flowing across the system and we are here interested
in the current fluctuations in the wire. Specifically, let $Q$ be the net
particle transfer from the left reservoir into the system during a time
interval $\tau$.  For a fixed $\tau$ we want to obtain the distribution
$P(Q,\tau)$ of $Q$, in the steady state of the system.  It is useful to
define the joint probability distribution function $R(Q, \mathcal{C},
\tau)$ for $Q$ number of particles transported and for the system to be in
state $\mathcal{C}$, given that at $\tau=0$ the system is in the steady
state.  Clearly $P(Q,\tau)= \sum_{\mathcal{C}} R(Q, \mathcal{C}, \tau)$. We
also define the characteristic functions $\tilde{R}(z,\mathcal{C},
\tau)=\sum_{-\infty}^{\infty}R(Q,\mathcal{C},\tau) z^Q$ and
$\tilde{P}(z,\tau)=\sum_{\mathcal{C}} \tilde{R} (z,\mathcal{C},\tau)$.  It
is then easy to obtain the following master equation \cite{DDR04}:
\begin{align}
  \f{d\tilde{R}(z,\mathcal{C},\tau)}{d\tau}=\sum_{\mathcal{C'}}\Bigl[&z
    \mathcal{W}_1(\mathcal{C},\mathcal{C'}) +
    \mathcal{W}_0(\mathcal{C},\mathcal{C'}) \notag
    \\ &+\f{1}{z}\mathcal{W}_{-1}(\mathcal{C},\mathcal{C'})\Bigr]\,\tilde{R}(z,\mathcal{C'},\tau).
\end{align} 
The general solution of this equation for arbitrary $L$ is difficult but
for $L=1$ an explicit solution can be obtained for $\tilde{R}
(z,\mathcal{C'},\tau)$ and $\tilde{P}(z,\tau)$. We will here first discuss
a special case $\alpha=\beta=\gamma=\delta$ for which $\tilde{P}(z,\tau)$
can be inverted explicitly.  The choice of steady state initial conditions
gives the solution: $ P(Q,\tau)= ({e^{-2\a \tau}}/{2})[I_{2Q-1}(2\a \tau) +
  2 I_{2Q}(2\a \tau) + I_{2Q+1}(2\a \tau)]$. In Fig.~(\ref{fig-1-part}) we
plot the exact distribution along with a direct simulation of the above
process with averaging over $5 \times 10^{8}$ realizations. As we can see
the direct simulation is accurate only for events with probabilities of
$O(10^{-8})$. Now we illustrate our algorithm using a biased dynamics.  We
consider biasing obtained by changing the boundary transition rates. We
denote the rates of the biased dynamics by $\a',\beta',\g', \delta'$ and
these are chosen such that $P_b(Q)$ has a peak in the required region. In
our simulation we consider a discrete-time implementation of SSEP. For
every realization of the process over a time $\tau$ (after throwing away
transients) the weight factor $W$ is dynamically evaluated. For instance,
every time a particle hops into the system from the left reservoir, $W$ is
incremented by $-\ln{(\alpha/\alpha')}$. In Fig.~(\ref{fig-1-part}) we see
the result of using our algorithm with two different biases. Using the same
number of realizations we are now able to find probabilities up to $
O(10^{-16})$ and the comparison with the exact result is excellent.

\begin{figure}
\includegraphics[width=3.25in]{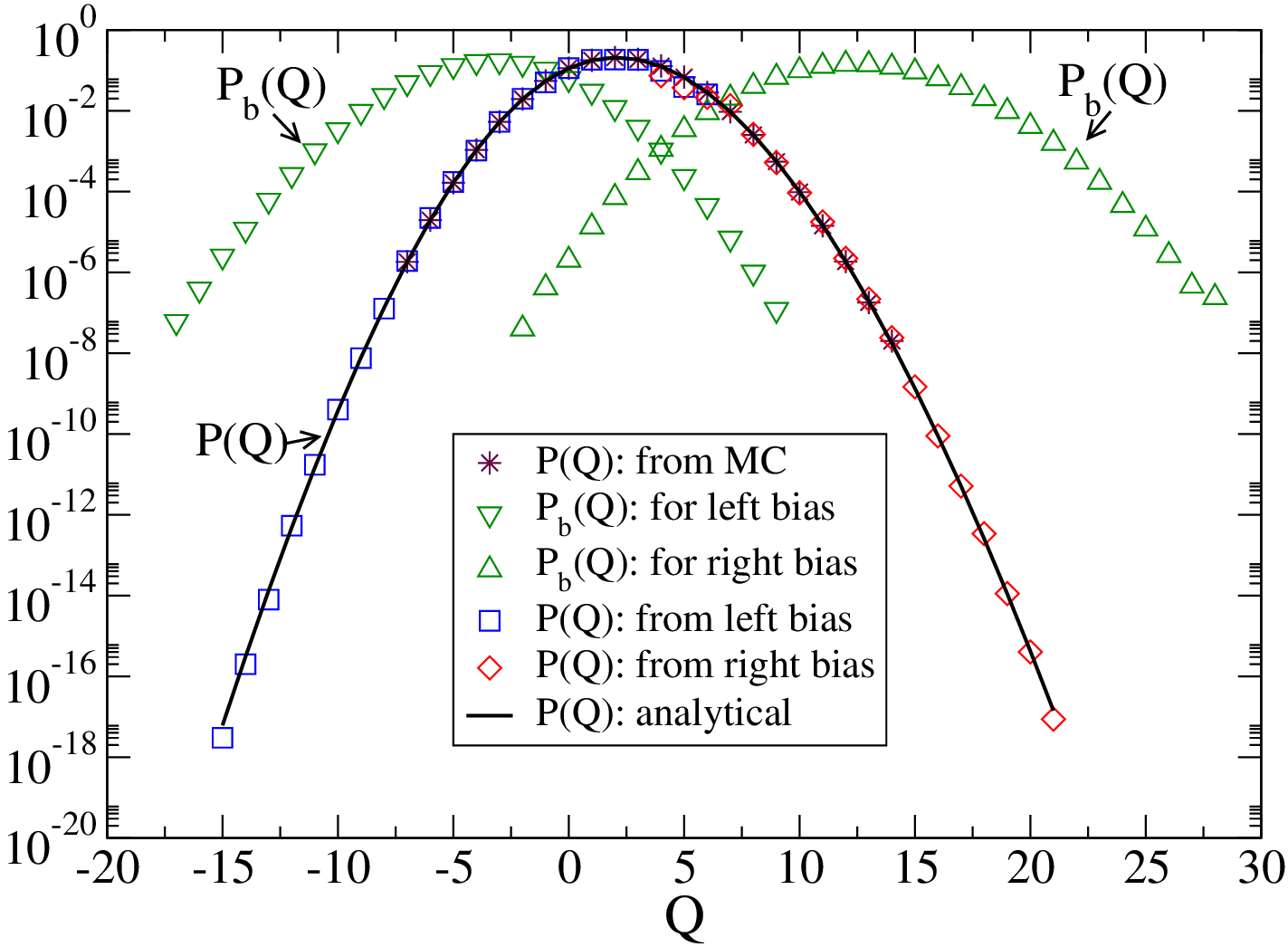}
\caption{(Color online) Plot of $P(Q)$ for $\tau=15$ for the three-site
  SSEP model with $\a=\bta=4.0,\g=\delta=2.0$. MC refers to direct
  Monte Carlo simulations.  For left (right) bias simulations, the
  particles in bulk hop to the left (right) with rate $4$ and to the right
  (left) with unit rate.  The boundary rates are kept unchanged.}
\label{fig-3-part}
\end{figure}

We next study the case with $L=3$ with rates chosen such that the system
reaches a nonequilibrium steady state with $\la Q \ra > 0$.  Finding
$\tilde{R}(z,\mathcal{C},\tau)$ analytically involves diagonalizing an $8
\times 8$ matrix. We do this numerically and after an inverse Laplace
transform find $P(Q,\tau)$. In Fig.~(\ref{fig-3-part}) we show the
numerical and direct simulation results for this case and also the results
obtained using the biased dynamics; in this case we consider a biased
dynamics with asymmetric bulk hopping rates. Again we find that the biasing
algorithm significantly improves the accuracy of finding probabilities of
rare events using the same number of realizations ($5 \times 10^8$).

\section{Heat conduction}

\begin{figure}
\includegraphics[width=3.25in]{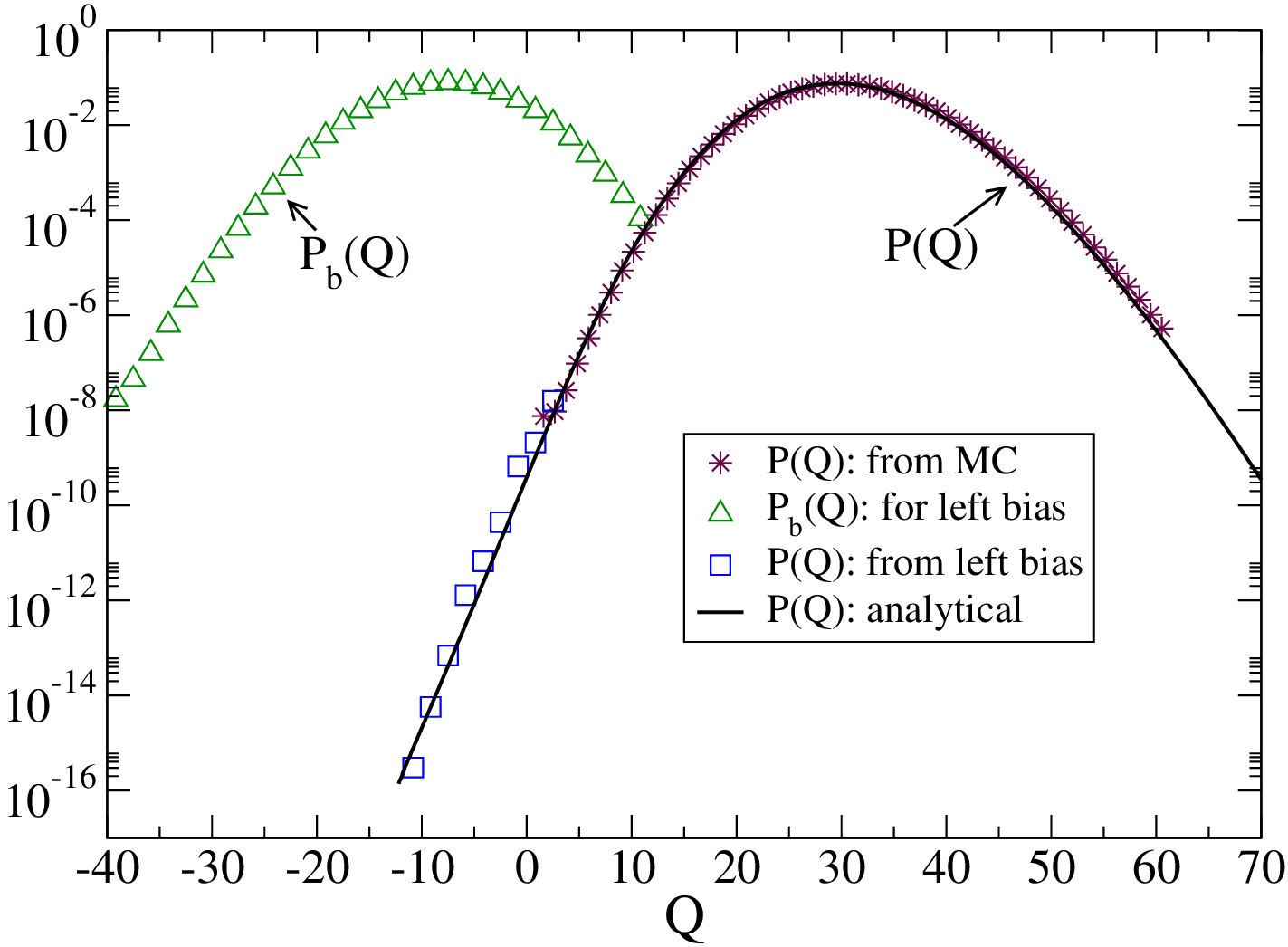}
\caption{(Color online) Plot of $P(Q)$ for $\tau=200$ for heat conduction
  across a single free particle with
  $\g_1=0.8,\g_2=0.2,T_1=1.1875,T_2=0.25$.  The parameters have been chosen
  to correspond to a region in parameter space where the fluctuation
  theorem is not satisfied \cite{visco06}.  MC refers to direct Monte Carlo
  simulations.  The left bias corresponds to $\g_1'=\g_1,\g_2'=\g_2/20,
  T_1'=T_1,T_2'=20 T_2$.}
\label{fig-visco}
\end{figure}

Next we consider the problem of heat conduction across a system connected
to heat reservoirs modeled by Langevin white-noise reservoirs. Here we are
interested in the distribution of the net heat transfer $Q$ from the left
bath into the system over time $\tau$.  First let us consider the simple
example of a single Brownian particle connected to two baths at
temperatures $T_1$ and $T_2$. This model was studied recently by
Visco~\cite{visco06} who obtained an exact expression for the
characteristic function of $Q$. The equation of motion for the system is
given by:
\begin{equation} 
\dot{v}=-(\gamma_1+\gamma_2)v +
\sqrt{2D_1}\,\eta_1 + \sqrt{2D_2}\,\eta_2 
\label{visco1}
\end{equation}
where $\eta_{1,2}$ are Gaussian delta-correlated noises with zero mean and
unit variance , thus $\la \eta_i(t)\eta_j(t')\ra = \delta_{ij}\delta(t-t')$
and $D_i=\g_iT_i$.  The heat flow from the left bath into the system in
time $\tau$ is given by $Q(\tau)=\int_{0}^{\tau}(-\gamma_1 v^2 +
\sqrt{2D_1} \eta_1 v)\, dt$. For the single Brownian particle in this
problem it is sufficient to specify the state by the velocity $v(t)$ alone.
If we choose $T_1 > T_2$ then $P(Q,\tau)$ will have a peak at $Q >0$.  It
is clear that to use the biasing algorithm to compute probabilities of rare
events with $Q <0$ we can choose a biased dynamics with temperatures of
left and right reservoirs taken to be $T_1'$ and $T_2'$ with $T_1'< T_2'$.
The calculation of the weight factor $W$ is somewhat tricky since computing
$\mathcal{P}[v(t)]$ from $\mathcal{P}[\eta_1(t),\eta_2(t)]$ is non-trivial.
Also one cannot eliminate $\eta_1$ to express $Q$ as a functional of only
the path $v$.  To get around this problem we note the following mapping of
the single-particle system to the over-damped dynamics of two coupled
oscillators \cite{wijland} given by the equations of motion: $\dot{x}_1 =
-\g_1(x_1-x_2) + \sqrt{2D_1}\eta_1~,~ \dot{x}_2 = -\g_2(x_2-x_1) -
\sqrt{2D_2}\eta_2$.  The variable $x_1-x_2=x_{12}$ satisfies the same
equation as $v$ in Eq.~(\ref{visco1}).  Thus with the same definition for
$Q$ as given earlier we can use the above equations for $x_1$ and $x_2$ to
find $P(Q,\tau)$.  In this case we do not have the problem as earlier and
both $Q$ and $W$ can be readily expressed in terms of $\{x_1,x_2\}$.  Let
us denote by $\gamma_i', T_i', D_i'$ the parameters of the biased
system. Also let $\eta_{1,2}'$ be the noise realizations in the biased
process that result in the same path $\{x_1,x_2\}$ as produced by
$\eta_{1,2}$ for the original process. Choosing $D_i=D_i'$ for $i=1,2$ it
can be shown that:
\begin{equation}
W=\int_0^\tau dt ~[ (\eta_1^2/2 + \eta_2^2/2) - (\eta_1'^2/2 + \eta_2'^2/2)
].  \label{weq}
\end{equation}
Using the equations of motion we can express
$\eta_{1,2},\eta'_{1,2}$ in terms of the phase-space variables and this
gives:
\begin{align*}
 W&=\f{1}{4D_1}\int_0^\tau dt[2(\g_1-\g_1')\dot{x}_1x_{12}+(\g_1^2-\g_1'^2)
   x_{12}^2]  \\ &+\f{1}{4D_2}\int_0^\tau dt[2(\g_2-\g_2')\dot{x}_2
   x_{12}+(\g_2^2-\g_2^2) x_{12}^2], \\ 
Q&=\int_0^\tau dt \dot{x}_1 x_{12}. 
\end{align*}
 Thus $W$ and $Q$ are easily evaluated in the simulation using the biased
 dynamics. In Fig.~(\ref{fig-visco}) we show results for $P(Q,\tau)$ obtained
 both directly and using the biased dynamics. Again we see that for the
 same number of realizations ($10^9$) one can obtain probabilities
 about $10^8$ times smaller than using direct simulations. The comparison
 with the numerical results obtained from the exact expression for $\la
 e^{-\lambda Q} \ra $ \cite{visco06} also shows the accuracy of the
 algorithm.

\begin{figure}
\vspace{0.75cm}
\includegraphics[width=3.25in]{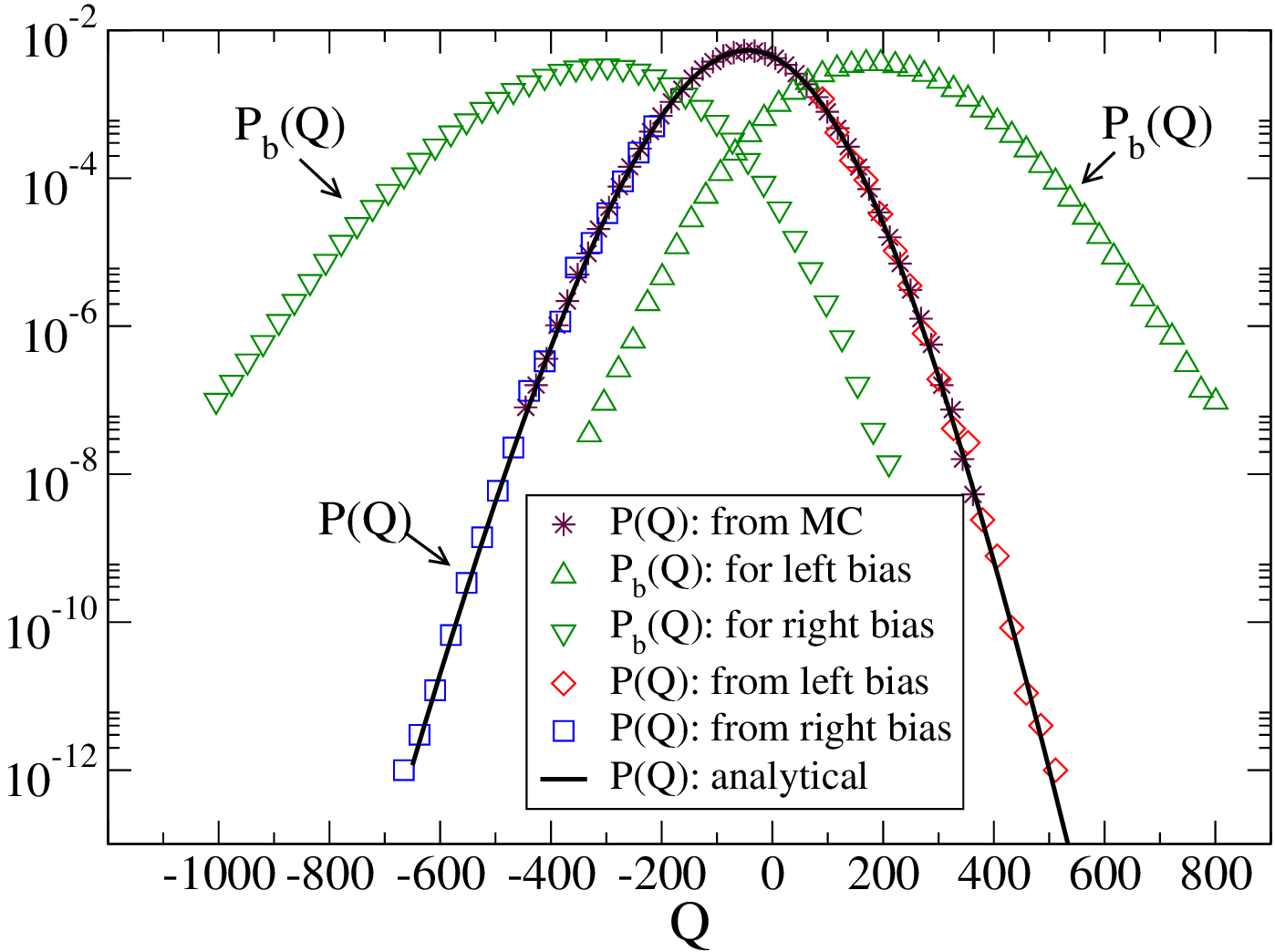}
\caption{(Color online) Plot of $P(Q)$ for $\tau=100$ for heat conduction
  across two particles connected by a harmonic spring with unit spring
  constant and $\g_1=\g_2=\sqrt{2},T_1=10,T_2=12$.  MC refers to direct
  Monte Carlo simulations.  The left bias corresponds to
  $\g_1'=\g_1,\g_2'=\g_2/2, T_1'=T_1,T_2'=2 T_2$ and right bias to
  $\g_1'=\g_1/2, \g_2'=\g_2, T_1'=2 T_1, T_2'=T_2$.}
\label{two-part}
\end{figure}

It is easy to apply the algorithm to more complicated cases. For example
consider a one-dimensional chain of $L$ particles connected to heat
reservoirs at the two ends with the following equations of motion:
\begin{align}
m_l \dot{v}_l &= f_l +\delta_{l,1}[-\gamma_1 v_1+\sqrt{2 D_1}\, \eta_1]
\notag \\ &+ \delta_{l,L} [ -\gamma_2 v_L + \sqrt{2 D_2}\,
  \eta_2],~~~l=1,2,\dots,N~, \label{geqm}
\end{align}
where $f_l=-\partial_{x_l}U$ and $U(\{x_l\})$ is the potential energy of
the system.  The net heat transfer from the left bath into the system is
given by $Q= \int_0^\tau (-\gamma_1 v_1^2 + \sqrt{2D_1} \eta_1 v_1)$ and
using Eqs.~(\ref{geqm}) this can be expressed in terms of $\{x_l,v_l\}$ as
$Q= \int_0^\tau dt v_1 (m_1 \dot{v}_1 -f_1) $.  To apply our algorithm we
consider a biased dynamics where the Hamiltonian evolution is unchanged
while the bath dynamics has new parameters $\gamma'_1,\g'_2, T'_1,T'_2$
which are chosen so that $P_b(Q)$ has a peak in the required
region. Choosing $D_i'=D_i$ we again find $W$ by using Eqs.~(\ref{geqm}) in
Eq.~(\ref{weq}), as for the single particle case.  Thus both $Q$ and $W$
can be expressed in terms of the path and so are readily evaluated for
every realization of the biased dynamics.

As an example we study the case $L=2$ with $U=(x_1-x_2)^2/2$ and with
$m_1=m_2=1$. For the special parameters $\g_1=\g_2=\sqrt{2}$ we use the
results in Ref.~\cite{SD07} to obtain $\la e^{-\lambda Q} \ra \sim
e^{\mu(\lambda) \tau}$ with $
\mu(\lambda)=\sqrt{2}\bigl\{1-[1+{\beta_1^{-1} \beta_2^{-1} \lambda (\Delta
    \beta -\lambda)}]^{1/6} \bigr\} $~.  This can be inverted to
numerically compute $P(Q,\tau)$ at large $\tau$.  In Fig.~(\ref{two-part})
we give the comparison between the analytical distribution and that
obtained by the biasing method.

\section{Conclusion} 

In conclusion, we have presented an algorithm for computing the
probabilities of rare events in various nonequilibrium processes. The
algorithm is an application of importance sampling and consists in using a
biased dynamics to generate the required rare events. This algorithm is
straightforward to understand and also to implement. The error in the
estimate of $P(Q,\tau)$ is $\approx \la
e^{-2W}\delta_{Q,Q_\bx}\ra_b^{1/2}/[MP_b(Q)]^{1/2}$. In the systems that we
have studied we find that the error can be made small by choosing the
biased dynamics carefully.  We have applied the algorithm to two different
models of particle and heat transport and shown that in both cases it gives
excellent results.  We note, however, that, in general, the fluctuations in
$W$ grow with $\tau$ and with the system size, hence the errors are large
and finding an appropriate biased dynamics is not always easy. Further work
is necessary for improving the efficiency of the algorithm for general
systems.

\begin{acknowledgments}
We thank S. R. S. Varadhan for useful discussion and suggestions.  
\end{acknowledgments}

\end{document}